\documentclass[12pt]{iopart}
\usepackage{graphicx}

\newcommand{\be}{\begin{eqnarray}}
\newcommand{\ee}{\end{eqnarray}}
\newcommand{\no}{\nonumber}

\begin{document}
\title{Ensemble Inequivalence and the Spin-Glass Transition}
\author{Zsolt Bertalan\footnote{zsolt@stat.phys.titech.ac.jp} 
and Kazutaka Takahashi\footnote{ktaka@stat.phys.titech.ac.jp}}
\address{Department of Physics, 
Tokyo Institute of Technology, Tokyo 152-8551, Japan}
\begin{abstract}
We report on the ensemble inequivalence 
in a many-body spin-glass model with integer spin. 
The spin-glass phase transition is of first order 
for certain values of the crystal field strength and 
is dependent whether it was derived 
in the microcanonical or the canonical ensemble. 
In the limit of infinitely many-body interactions, 
the model is the integer-spin equivalent of the random-energy model, 
and is solved exactly.
We also derive the integer-spin equivalent of 
the de Almeida-Thouless line.
\end{abstract}

\maketitle

\section{Introduction}

The circumstance that two statistical ensembles lead to 
different results for some systems, called ensemble inequivalence, 
is one of the most perplexing phenomena in statistical physics. 
It was known for 
gravitational systems \cite{antonov, LBW,LB,thirring,HT}
long before the application to spin systems ten years ago \cite{BMR}. 
The re-discovery of ensemble inequivalence 
sparked renewed interest and brought about many studies 
(for reviews see \cite{CDR,DRAW,mukamel,DRC,chavanis}). 
A prominent feature of ensemble inequivalence is 
the emergence of negative specific heat in the microcanonical ensemble. 
In the canonical ensemble, specific heat is always positive, 
since the free energy is always a concave function. 
On the other hand, the microcanonical entropy as a function of 
the internal energy, around a first order transition can have 
a stable convex intruder \cite{gross}. 
Another feature of ensemble inequivalence is the occurrence of 
mixed phases in the microcanonical phase diagram, 
when the temperature and another, possibly external, 
control parameter are chosen as axes. 
This is the direct result of a convex intruder in the entropy 
as a function of energy, since then two distinct phases 
can exist at the same temperature. 
For more intriguing properties of this phenomenon 
we refer the reader to the literature, e.g. \cite{CDR}. 

The best known occurrence of ensemble inequivalence is 
in long-range interacting systems which have 
a first-order phase transition \cite{ETT1,ETT2,BoBa}, 
but similar properties also occur in driven systems 
with local dynamics \cite{LM,LCM}. 
An interaction is said to be of long range when it decays slower 
than $1/r^{\alpha}$, where $r$ is the spatial distance between 
interacting objects and the value of $\alpha$ is smaller than 
the system's dimension, 
or when the system size is of comparable order 
as the range of the interaction. 
In long-range interacting systems, the energy is not additive, 
i.e. if two identical sub-systems with energy $E$ are brought 
into contact, the total energy of the resulting system is not $2E$, 
because interface interactions cannot be neglected.

Such systems appear in virtually every branch of physics, 
from atomic physics \cite{Schmidt} 
to spin systems \cite{BMR,IC, MRS,BMR2,BDMR,BKMN,BN} 
and gravitational systems (see \cite{chavanis} and references therein). 
Recently, the effect of randomness on ensemble inequivalence 
was investigated using many-body interacting 
spin models in \cite{BKMN,BN}. 
So far, however, no inequivalence was found 
for the spin-glass (SG) transition.  

The results of the theory of SGs is of interest 
for many areas of science. 
Fields like statistical mechanics, information processing, 
image restoration and neural networks take advantage of the methods 
first used to describe SGs, 
see \cite{MPV,Nishimori,MM} and references therein. 
Therefore, it will be an important task to study 
how the ensemble inequivalence appears in SG systems.

In this paper, we investigate a system which shows 
a first-order transition between paramagnetic and SG phases: 
the infinite-range, many-body SG model 
with integer spin, $S=1$. 
This model is the mean-field variant of the Blume-Emery-Griffiths 
model \cite{BEG}, which was introduced to describe mixtures of liquid He3-He4, 
but with quenched, random interactions. 
The case of random interactions was first studied for two-body interactions 
in \cite{GS} by using the simple replica symmetric ansatz. 

Infinite-range models have the advantage that they are exact 
in the mean-field ansatz and thus yield valuable limiting cases 
of more realistic models. 
Apart from interest for investigations of ensemble inequivalence, 
our model is another example of the (admittedly, classical) Bose-glass phase, 
which appears in systems of Helium in porous media \cite{FWGF}. 
Moreover, taking the limit of infinitely many-body interactions yields 
for our model an analytically accessible scenario and a generalization 
of the random energy model (REM) \cite{Derrida1, Derrida2,GM, OFY}.

The organization of the paper is as follows. 
In \sref{model}, we introduce the model under investigation, 
and state the results for the thermodynamical potentials. 
In \sref{REM}, we draw the analytically accessible phase diagram 
of the REM, 
corresponding to the limit of infinitely-many-body interactions. 
We also study three-body interacting case in \sref{sec:p3}
and $p$-body case with $p>3$ in \sref{sec:pgt3}.
To clarify the analytical result at $p\to\infty$, 
we investigate numerically how the infinite limit is attained. 
We conclude our investigations in \sref{conclusio}.
In the appendices, we show some technical details of the calculations, 
in \ref{details} we describe the derivation of 
the replicated thermodynamic potentials, while in \ref{AT} 
we derive the de Almeida-Thouless (AT) condition \cite{AT,BT} 
for the many-body integer-spin case.

\section{Model}\label{model}

We investigate the properties of a system of $N$ spins, 
where each spin can take the values $+1, 0, -1$, 
and each interacts with every other spin through quenched, 
random $p$-body interactions. 
It is given by the Hamiltonian
\be
 H = -\sum_{i_1<i_2<\cdots <i_p}J_{i_1i_2\cdots i_p}
 S_{i_1}S_{i_2}\cdots S_{i_p}
 +D\sum_{i=1}^N(S_i)^2.
 \label{hamilton}
\ee
The role of the external field is played by the so-called crystalline anisotropy, 
whose strength is measured by the parameter $D$. 
The interactions $J_{i_1i_2\cdots i_p}$ are quenched random numbers 
obeying the Gaussian probability distribution 
\be
 {\rm Prob}(J_{i_1\cdots i_p})
 = \left(\frac{N^{p-1}}{\pi p!}\right)^{1/2}
 \exp\left\{
 -\frac{N^{p-1}}{p!}(J_{i_1\cdots i_p})^2\right\}.
\ee
The limit $D\to-\infty$ corresponds to the Ising model, i.e. 
only the spin states $S_i=\pm 1$ contribute to the energy. 

The corresponding pure system, namely when the interactions are not 
random and are all equal, is known as 
the Blume-Emery-Griffiths model~\cite{BEG}. 
It was shown in \cite{BMR} that the pure model 
with $p=2$ exhibits ensemble inequivalence 
in the region of the first-order phase transition. 
For technical reasons, explained below, 
we will treat only the many-body interacting case with $p>2$.

Using the standard methods of replica theory (see e.g. \cite{Nishimori}), 
we can calculate the thermodynamic potentials, 
the free energy in the canonical and 
the entropy in the microcanonical ensemble. 
We employ the one-step replica symmetry breaking 
(1RSB) \cite{Parisi1,Parisi2,Parisi3} ansatz
for the SG order parameter.
Here, we only give the results and 
the details are described in \ref{details}. 

In the canonical ensemble, 
the free energy density is obtained as a 
function of the order parameters as
\be
 \beta \phi &= 
 -\frac{\beta^2}{4}\chi^p+\hat{\chi}\chi
 +\frac{\beta^2}{4}(1-x)q^p-\frac{1-x}{2}\hat{q}q
 +\beta D\chi
 \no\\
 &\quad -\frac{1}{x}\ln
 \left\{
 \int Dz \left[
 1+ 2e^{\hat{\chi}-\hat{q}/2}\cosh (\sqrt{\hat{q}}z)
 \right]^x\right\},\label{phi}
\ee
while the entropy density in the microcanonical ensemble is from 
\be
 \sigma = -&\frac{(\epsilon-D\chi)^2}{\chi^p-(1-x)q^p}
 -\hat{\chi}\chi
 +\frac{1-x}{2}\hat{q}q\no\\ & +\frac{1}{x}\ln\left\{\int Dz
 \left[1+2e^{\hat{\chi}-\hat{q}/2}
 \cosh (\sqrt{\hat{q}}z)\right]^x
 \right\}, \label{sig}
\ee
where $\beta=1/T$ is the inverse temperature, 
$\epsilon=E/N$ is the energy per spin 
and $Dz=dz\exp(-z^2/2)/\sqrt{2\pi}$.
The symbols $\chi$, $q$ and $x$ 
are variables parametrizing the 1RSB solution. Here, $q$ is the spin glass order parameter, $\chi$ is the quadrupole moment and 
$x$ is the Parisi RSB parameter. The order parameters have the following physical meaning:
\be
	q=[\langle S_i^{\alpha}S_i^{\beta}\rangle]\quad\textrm{and }\chi=[\langle (S_i^{\alpha})^2\rangle],
\ee
where $[\ldots]$ and $\langle\ldots\rangle$ represent the configurational and thermal average, respectively. Therefore, $q$ can be thought to measure the correlation of spins in different replicas and $\chi$ measures the fraction of spins in the Ising states $S_i\in{\pm 1}$ within a single replica. The parameter $x$ has no clear physical meaning, but can be thought to measure the degree of RSB in the system.
	
The variables with the hat symbol, $\hat q$ and $\hat \chi$, are the Fourier modes of $q$ and $\chi$, respectively. The modes are related to the physical quantities through saddle-point equations specified in \ref{details}.
We can easily see that 
the generalized thermodynamic potentials
$\phi$ and $\sigma$ are formally related via 
$\beta \phi=\beta\epsilon -\sigma$, when the caloric relation 
\be
 \beta = -\frac{2(\epsilon-D\chi)}{\chi^p-(1-x)q^p}
 \label{caloric}
\ee
is used.

The stable branches of the free energy 
are determined by the condition 
\be
 f(\beta)=\min_{\chi}\max_{q,x}\phi(\chi,q,x,\beta),
\ee 
and of the entropy by
\be
 s(\epsilon)=\max_{\chi}\min_{q,x}\sigma(\chi,q,x,\epsilon).
\ee
 These extremization conditions give the saddle-point equations
\be
 & \chi=\int D'z\frac{2e^{\hat{\chi}-\hat{q}/2}
 \cosh(\sqrt{\hat{q}}z)}
 {1+2e^{\hat{\chi}-\hat{q}/2}\cosh(\sqrt{\hat{q}}z)}, 
 \label{spe1} \\
  & q= \int D'z\left\{\frac{2e^{\hat{\chi}-\hat{q}/2}
 \sinh(\sqrt{\hat{q}}z)}
 {1+2e^{\hat{\chi}-\hat{q}/2}\cosh(\sqrt{\hat{q}}z)}\right\}^2  
 \label{spe2}
 \ee
 and 
 \be
  \left(\frac{1}{2}q\hat q -\frac{\beta^2}{4}q^p\right)x^2  = & x \int D'z \ln\left(1+2e^{\hat{\chi}-\hat{q}/2} \cosh(\sqrt{\hat{q}}z)\right) \\ & \quad -\ln\left\{\int Dz \left[1+2e^{\hat{\chi}-\hat{q}/2}\cosh(\sqrt{\hat{q}}z)\right]^x\right\},
\ee
 where
\be
 \int D'z\left(\cdots\right)
 = \frac{\int Dz \left\{1+2e^{\hat{\chi}-\hat{q}/2}
 \cosh(\sqrt{\hat{q}}z)\right\}^x
 \left(\cdots\right)}
 {\int Dz \left\{1+2e^{\hat{\chi}-\hat{q}/2}
 \cosh(\sqrt{\hat{q}}z)\right\}^x}.
 \label{Dz}
\ee

As stated, the saddle-point equations lead to 
multiple branches of $\sigma$ and $\phi$, 
which all correspond to various stable, meta-stable and unstable phases. 
Ensemble inequivalence occurs when there is no one-to-one correspondence 
of the stability of the branches of the thermodynamic potentials. 
It may happen, for example, that a stable branch of the entropy 
corresponds to a meta-stable, or even unstable branch of the free energy.
That is a macrostate realized in
the microcanonical ensemble, might not exist in the canonical ensemble.
This scenario appears, however, only for first-order phase 
transitions \cite{CDR}, since at second-order transitions 
there are no meta-stable solutions of the saddle-point equations.

The final remark to be made in this section is 
on the validity of the 1RSB ansatz.
In the standard $p$-body Ising SG with $S_i=\pm 1$, 
for $p>2$ the replica-symmetric solution of the SG phase does not show up 
and the replica symmetry is already broken
just below the SG transition temperature, justifying the 1RSB-ansatz.
At low temperatures, the 1RSB solution becomes unstable and 
the SG phase is described by 
the full-step replica symmetry breaking solution ~\cite{Gardner,NW}.
We expect that a similar scenario holds in the present integer-spin case.
In the following, we will show that this is indeed the case.
However, for $p=2$, the SG phase is described by 
the full-step replica symmetry breaking solution 
and the 1RSB solution does not appear.
That is the reason why we consider only $p>2$. 
For a canonical analysis with $p=2$, see \cite{GS}.

\section{Infinitely-many-body interactions ($p\to\infty$)}\label{REM}

In the limit $p=\infty$, our model \eref{hamilton} is 
the spin-1 equivalent of the REM \cite{Derrida1,Derrida2,GM}. 
The saddle-point equations \eref{spe1} and \eref{spe2} 
can be solved analytically and have three simple solutions.
These yield three respective branches of the free energy and the entropy, 
which we identify as paramagnetic (P), Ising paramagnetic (IP), 
and SG phases. These phases are characterized through the order parameters as
\be
\begin{array}{lll}
 {\rm P:} & q=0, & \chi < 1, \\
 {\rm IP:} & q=0, & \chi = 1, \\
 {\rm SG:} & q=1, & \chi = 1, \\
\end{array}\no
\ee
in both ensembles. 
It is thus not a-priory clear where ensemble inequivalence should occur 
on the level of macrostates. 
The aims of the present section is to clarify this point. 
This will be done by comparing 
the canonical and microcanonical phase diagrams.

\subsection{Canonical ensemble}

The saddle-point equations always allow the solution $q=0$.
In that case, we get a self-consistent equation for $\chi$ as 
\be
 \chi = \frac{2\exp\left(-\beta D+\frac{\beta^2}{4}p\chi^{p-1}\right)}
 {1+2\exp\left(-\beta D+\frac{\beta^2}{4}p\chi^{p-1}\right)}.
\ee
In the limit $p\to\infty$, there are two possibilities, 
$\chi<1$ ($\chi^{p}\to 0$) and $\chi=1$, 
corresponding to the P and IP phases respectively. 

In the P phase, the order parameter 
\be
 \chi = \frac{2e^{-\beta D}}{1+2e^{-\beta D}},
\ee 
 corresponds to the fraction of spins that take the value $\pm 1$.
 The free energy in this phase is given by 
\be
 f_{\rm P}=-\frac{1}{\beta}\ln\left(1+2e^{-\beta D}\right).
\ee 
In the IP phase, the zero-spin state plays no role and 
each spin takes the value $\pm 1$ only, so that $\chi=1$.
The free energy here is 
\be
 f_{\rm IP} = D-\frac{1}{\beta}\ln 2-\frac{\beta}{4},
\ee
which coincides with the standard Ising result as expected, 
except for a trivial shift by $D$.
At temperatures lower than the critical temperature defined by 
\be
 T_{\rm c} = \frac{1}{2\sqrt{\ln 2}}, 
\ee
the IP state becomes unphysical, and the spin-glass phase has to take over. 
For any $q<1$, 
the saddle-point equations yield the solution $q=0$ and 
the SG phase appears only at $q=1$. 
Here, we have also $\chi =1$ from the requirement $\chi\ge q$.
Then, the free energy takes the form
\be
 f = D-\frac{\beta}{4}x-\frac{1}{\beta x}\ln 2.
\ee
The parameter $x$ is determined by the extremal condition 
$\partial f/\partial x=0$ as
\be
 x = \frac{T}{T_{\rm c}}=2T\sqrt{\ln 2}.
\ee
 We obtain the SG phase solution as
\be
 f_{\rm SG} = D-\sqrt{\ln 2},
\ee
where the zero spin state is irrelevant and 
the result is essentially the same as in the standard REM.  

\begin{figure}[htb]
\includegraphics[width=\columnwidth]{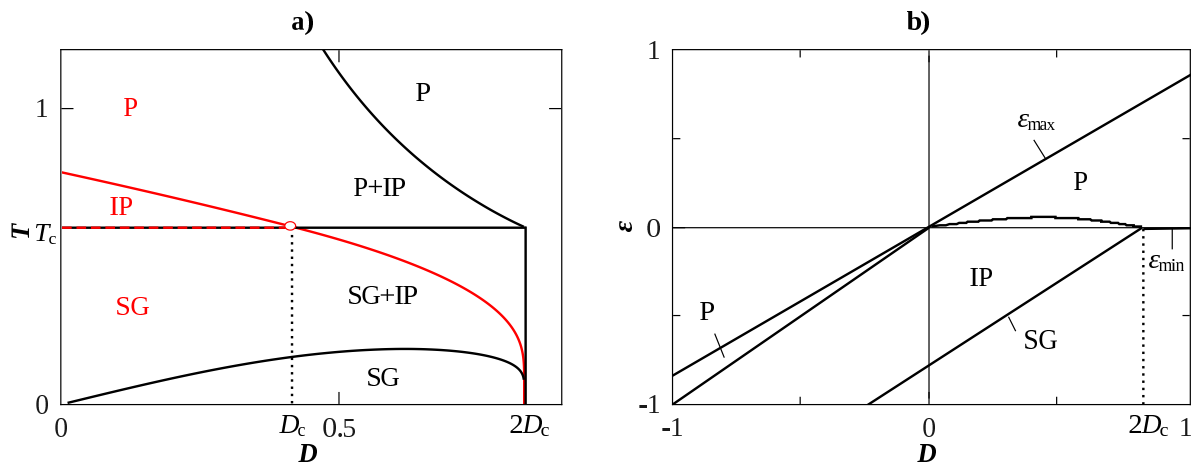}
\caption{a) Phase diagram of the integer-spin REM in the $(D,T)$-plane. 
The canonical phase boundaries are drawn in red, 
and the microcanonical phase boundaries are overlaid in black for comparison. 
The canonical (microcanonical) phases are marked in red (black) letters. The red circle indicates the tricritical point and the second order IP-SG transitions line is the red dashed line.
b) Microcanonical phase diagram in the $(D,\epsilon)$-plane.}
\label{phase-ct}
\end{figure}

The phase boundaries can be easily obtained analytically 
by equating the free energies of two respective phases. 
We get each phase boundary as follows.
\be
 \mbox{P-IP}:& 1+2e^{-\beta D} = 2e^{-\beta D+\beta^2/4}, \\
 \mbox{IP-SG}:& T=T_{\rm c}, \\
 \mbox{P-SG}:& 1+2e^{-\beta D} = 2e^{-\beta D+\beta\sqrt{\ln 2}}.
\ee
The resulting canonical phase diagram is drawn in \fref{phase-ct}a) 
with red lines. 
The SG phase exists at low temperatures for $D<2D_{\rm c}$
where 
\be
 D_{\rm c}=\frac{\sqrt{\ln 2}}{2}.
\ee
The P-SG transition for $D_{\rm c}<D<2D_{\rm c}$ is of first order. 
At $D=D_{\rm c}$, there is a tricritical point 
where the P, IP and SG phases meet. 
For $D<D_{\rm c}$, there is a first-order P-IP transition, 
whose transition temperature increases with decreasing $D$. 
The second-order IP-SG transition is at the critical 
temperature $T_{\rm c}$ for any $D<D_{\rm c}$, thus 
regardless of the crystal field strength. 
As $D\to -\infty$, the P-IP transition temperature diverges 
and we recover the standard REM result with only the IP and SG phases present.

\subsection{Microcanonical ensemble}

\begin{figure}[htb]
\begin{center}
\includegraphics[width=0.8\columnwidth]{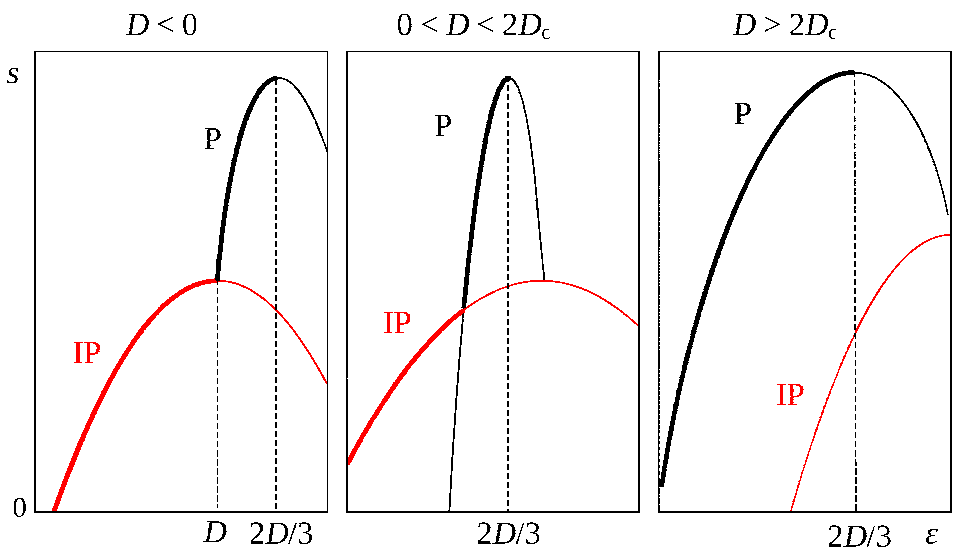}
\caption{Entropy density $s(\epsilon)$ in the IP (red) and P (black) phases.
The equilibrium entropy follows bold parts. 
The leftmost panel shows the case for $D<0$, as can be seen, 
the P phase reaches all temperatures in the interval $[0,\infty)$, 
while the IP phase exists in the range $[T_{\rm c},\infty)$ 
where both phases coexist. 
The middle panel shows a situation where there is a P-IP transition 
at finite temperature is realized, 
while the rightmost panel shows a situation where the IP phase is unstable.
}
\label{ms}
\end{center}
\end{figure}

Similarly to the canonical case, 
the solutions of the microcanonical entropy 
can be calculated explicitly and we have
\be
 s_{\rm IP} =  \ln 2-(\epsilon-D)^2, \\
 s_{\rm P} = \frac{\epsilon}{D}\ln 2
 -\frac{\epsilon}{D}\ln\frac{\epsilon}{D}
 -\left(1-\frac{\epsilon}{D}\right)
 \ln\left(1-\frac{\epsilon}{D}\right), \\
 s_{\rm SG} = 0.
\ee
The phase diagram is determined by comparing these results, 
but in contrast to the canonical case, the higher entropy wins.
Physics requires furthermore that the entropy is positive.
We plot the entropy densities in the P (black) and IP (red) phases in \fref{ms}.
When the entropy reaches zero, the system freezes to the ground state
and the SG phase is realized (not shown).
In each phase, the energy range is specified by
\be
 \mbox{IP}:& D-\sqrt{\ln 2}<\epsilon<D, \\
 \mbox{P}:& 0<\epsilon<\frac{2}{3}D \ (D>0)
 \quad\mbox{and}\quad
 D<\epsilon<\frac{2}{3}D \ (D<0), \\
 \mbox{SG}:& \epsilon=D-\sqrt{\ln 2}.
\ee
We note furthermore that in \fref{ms} the region where 
the respective entropies are stable are drawn in thick lines. 

The microcanonical phase diagram in the $(D,\epsilon)$-plane 
is shown in \fref{phase-ct}b).
In this case, the values that the energy density 
for each phase can take are limited, 
denoted by $\epsilon_{\rm min}$ and $\epsilon_{\rm max}$ respectively. 
We also see that the SG phase lies on a single line 
in the phase diagram, because it exists 
only for one value of the energy for a fixed $D$. 
For $D<2D_{\rm c}$ this is the $\epsilon_{\rm min}$-line. 
Therefore, the temperature $1/T=ds/d\epsilon$ is not well defined in the SG-phase 
and has to be extrapolated from finite $p$-phase diagrams.

The ensemble inequivalence becomes apparent 
when we change the dependent variable from $\epsilon$ to $T$ 
according to relation \eref{caloric} and 
thus transform the microcanonical phase diagram 
from the $(D,\epsilon)$-plane to the $(D,T)$-plane. 
The result is shown in figure~\ref{phase-ct}a) in black.
We see that it is qualitatively different from the canonical result 
shown in the same figure in black.
The IP-SG transition is of second order and 
that phase boundary coincides in both ensembles.
The other boundaries are totally different from each other.
With reference to \fref{ms} we see that 
there is no one-to-one correspondence between $T$ and $\epsilon$ 
for crystal-field strengths $D<2D_{\rm c}$.
There are regions such that two values of $\epsilon$ correspond
to the same temperature.
As a result, we have two coexistence phases P+IP and P+SG, as 
shown in \fref{phase-ct}a).

We find that the canonical and microcanonical phase diagrams are 
very different and it is hard to find common properties 
except the IP-SG boundary.
We presume that one of the reasons for this discrepancy 
is the peculiarity of the $p\to\infty$ limit \cite{BKMN,BN}.
Therefore, we will study the finite-$p$ case in the next section, 
to see how this deviations come about.

\section{Three-body interactions ($p=3$)}\label{sec:p3}

In this section, 
by solving the saddle-point equations with numerical help, 
we derive the phase diagrams 
of the Hamiltonian \eref{hamilton} 
for $p=3$.

\subsection{Phase diagram}

\begin{figure}[ht]
\includegraphics[width=\textwidth]{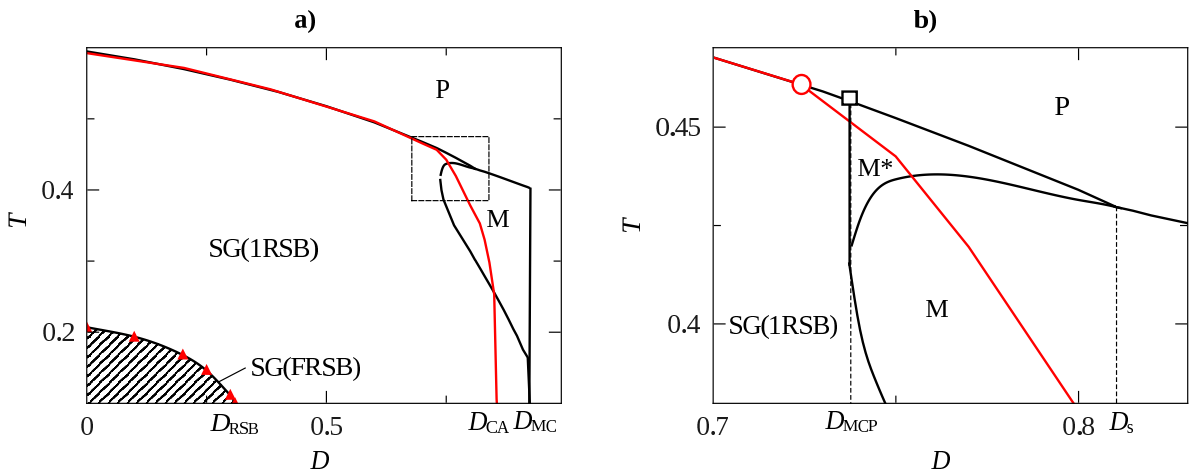} 
\caption{
a) Phase diagram in the $(D,T)$-plane with $p=3$. 
The microcanonical phase boundaries are shown in black and 
the canonical in red.
The black line with triangles is the AT line. 
b) Detail of a) denoted by the box with dashed edges. 
The red circle corresponds to the canonical multicritical point, where 
the P-SG transition goes from second order (left) to first order (right). 
The microcanonical multicritical point is at $D_{\rm MCP}$ and 
is symbolized by a black square.
}
\label{p3pd}
\end{figure}

In the canonical ensemble, 
there is only a phase transition from the paramagnetic phase 
to the SG phase.
We note that the IP phase 
appears only when the limit $p\to\infty$ is taken.
We draw the phase diagram in \fref{p3pd}a) in red. 
The transition is of second order for $D<0.724$ and 
of first order until the SG-phase terminates 
at $D=D_{\rm CA}\approx 0.84$.

\begin{figure}[ht]
\centering
\includegraphics[width=0.5\textwidth]{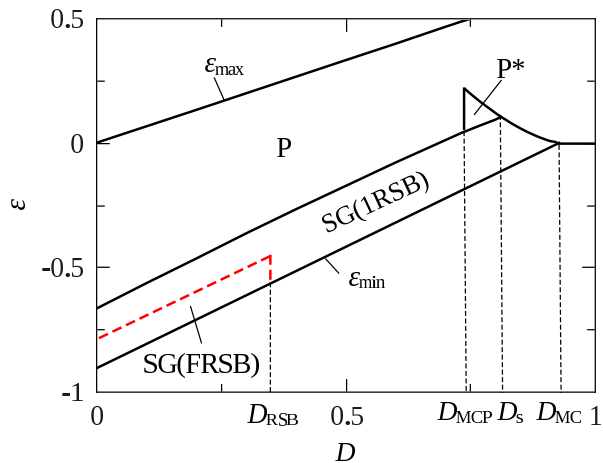} 
\caption{Microcanonical phase diagram in the $(D,\epsilon)$-plane. 
$\epsilon_{\rm max}$ and $\epsilon_{\rm min}$ are
the maximal and minimal attainable energies respectively. 
The red dashed line is the AT line. 
}
\label{p3pdb}
\end{figure}
\begin{figure}[ht]
\centering
\includegraphics[width=0.8\textwidth]{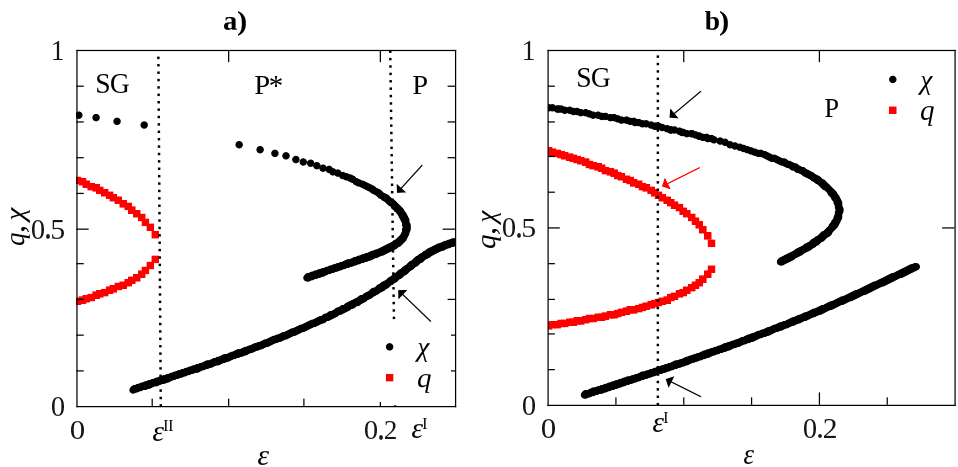} 
\caption{
Several branches of the saddle-point solutions 
of order parameters.  
a) For $D=0.745$, there are two successive transitions.
At $\epsilon>\epsilon^{\rm I}$, 
the system is in the P phase and the order parameter $\chi$ 
takes the lowest branch.
It changes discontinuously at $\epsilon=\epsilon^{\rm I}$
as indicated by the two black arrows. 
At $\epsilon^{\rm II}<\epsilon<\epsilon^{\rm I}$, 
the system is still in a paramagnetic state, 
but labeled by P$^{\rm *}$.
At $\epsilon=\epsilon^{\rm II}$ 
the SG order parameter (red squares) jumps from zero (not shown) 
to a finite value. 
b) The same as a) but at $D=0.83$. 
Here, the phase P$^{\rm *}$ is meta-stable throughout 
and thus does not show up. 
The P-SG transition is of first order, 
accompanied by a jump in $\chi$ and $q$, 
indicated by the black and red arrows.
} 
\label{p3ops}
\end{figure}

In the microcanonical ensemble, 
the phase diagram in the $(D,\epsilon)$-plane is shown in \fref{p3pdb}
and is summarized as follows. 
\begin{itemize}
\item{$D<D_{\rm MCP}\approx 0.737$} 

There is a second-order P-SG phase transition as in the canonical case. 

\item{$D_{\rm MCP}<D<D_{\rm s}\approx 0.81$}

In this case, as the energy is lowered, we first observe 
a first-order transition in the paramagnetic phase. 
The order parameter $\chi$ jumps
as depicted in figure \ref{p3ops}a). 
The jump in $\chi$ is at $\epsilon=\epsilon^{\rm I}$ 
and is indicated by the black arrows in the figure. 
We denote this paramagnetic phase P$^{\rm *}$. 
As the energy decreases further, there is a phase transition 
from P$^{\rm *}$ to SG at $\epsilon=\epsilon^{\rm II}$. 
The P$^{\rm *}$-SG transition is of second order 
because at the transition point the Parisi breaking parameter is $x=1$, 
guaranteeing a smooth change in the entropy. 

\item{$D_{\rm s}<D<D_{\rm MC}\approx 0.92$}

The phase P$^{\rm *}$ becomes meta-stable for all accessible energies 
and does not appear.
The P-SG transition is of first order. 
The order parameters are shown in figure \ref{p3ops}b) for $D=0.83$. 

\item{$D>D_{\rm MC}$}

In this case, there is no stable SG phase and 
we have only the P phase. 

\end{itemize}

When the energy is translated into temperature 
through the formula \eref{caloric},
we can draw the phase diagram in the $(D,T)$-plane as shown
in \fref{p3pd} (black lines).
Although the P-SG second-order transition line coincides,
the other boundaries do not match.
In the microcanonical calculation, the point where 
the transition changes from being second-order to first-order is  
$D=D_{\rm MCP}\approx 0.737$, 
while it is $D\approx 0.724$ in the canonical.
We also see that here the P$^{\rm *}$ phase does not show up separately. 
Instead there are two regions where phases coexist: 
M=P+SG and M$^{\rm *}$=P+SG+P$^{\rm *}$.
We will see in the next subsection how this result is obtained.

\subsection{Phase coexistence}

\begin{figure}[ht]
\includegraphics[width=0.8\textwidth]{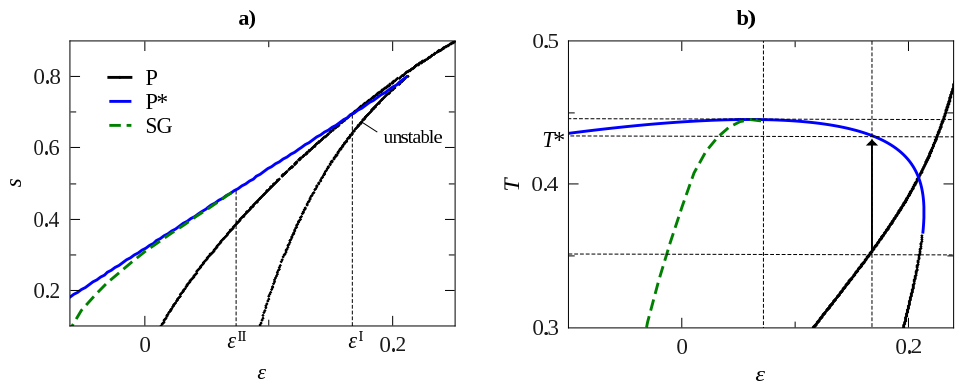} 
\caption{a) Branches of the entropy versus energy 
per spin for $p=3$ at $D=0.77$: P (black), 
P$^{\rm *}$ (blue) and SG (green dashed).
b) The corresponding caloric curve. 
The phase transitions are at $\epsilon^{\rm I}$ (P-P$^{\rm *})$ and 
at $\epsilon^{\rm II}($P$^{\rm *}$-SG).
} 
\label{p3es}
\end{figure}

Let us now explain shortly, how the mixed phases appear 
in the $(D,T)$-plane (\fref{p3pd}), 
but not in the $(D,\epsilon)$-plane (\fref{p3pdb}). 
We fix the external field at $D=0.77$, 
so that all phases show up and plot the entropy versus 
the energy in \fref{p3es}a). 
The corresponding caloric curve $T(\epsilon)$ is shown in \fref{p3es}b). 
At high energies $\epsilon>\epsilon^{\rm I}$, 
the system is in the P state (black). 
As the energy is lowered, the temperature decreases as well. 
At some point the system develops meta-stable and unstable solutions. 
The meta-stable state corresponds to P$^{\rm *}$ (blue), 
the unstable one has no physical meaning, 
but is plotted for the sake of completeness. 
At $\epsilon=\epsilon^{\rm I}$ the entropies of the P and P$^{\rm *}$ 
have the same value, but at the crossing point have different slopes,
\be
 \frac{ds_{\rm P}}{d\epsilon}(\epsilon^{\rm I})
 \neq \frac{ds_{{\rm P}^{\rm *}}}{d\epsilon}(\epsilon^{\rm I}).
\ee
Therefore, the temperature jumps at the transition point, 
$T^{\rm I}\to T^{\rm *} $, as indicated by an arrow in \fref{p3es}b). 
Noteworthy is now that as the energy is lowered 
the temperature $increases$, corresponding to negative specific heat 
of the P$^{\rm *}$ phase. 
At $\epsilon=\epsilon^{\rm II}$, 
the system allows for a SG solution (green dashed) with $q>0$. 
However, this SG-entropy connects smoothly to the P$^{\rm *}$-entropy, 
so that
\be
 \frac{ds_{\rm SG}}{d\epsilon}(\epsilon^{\rm II})
 = \frac{ds_{{\rm P}^{\rm *}}}{d\epsilon}(\epsilon^{\rm II}),
\ee
and there is no temperature jump. 
Thus, the transition is of second order, 
although the order parameter $q$ jumps as indicated by \fref{p3ops}. 
This is, as mentioned before, a
consequence of the replica trick and the 1RSB parameter $x$ 
guarantees a smooth transition.

Mixed phases appear when we choose a temperature between, 
say $T^{\rm II}$ and $T^{\rm *}$.
At such a temperature, we can find stable solutions 
for every phase, giving rise to the M$^{\rm *}$-mixed phase. 
The mixed phase M lies between $T^{\rm I}$ and $T^{\rm *}$, 
and we can not distinguish the SG and the P phases. 

\subsection{AT condition}

In \fref{p3pd} and \ref{p3pdb}, we plot the AT line, 
the boundary of the stability region.
Below that line, the 1RSB solution of the SG phase is unstable 
and the full-step replica symmetry breaking solution has to be considered. 
We derive the AT condition in \ref{AT}. 
As we have shown for the Ising system in \cite{BT},
the conditions in both ensembles 
are essentially the same and are translated via the caloric relation \eref{caloric}.
We confirm that this equivalence holds also in the present case.

\section{Many-body interactions ($p>3$)}
\label{sec:pgt3}

\begin{figure}[ht]
\includegraphics[width=\textwidth]{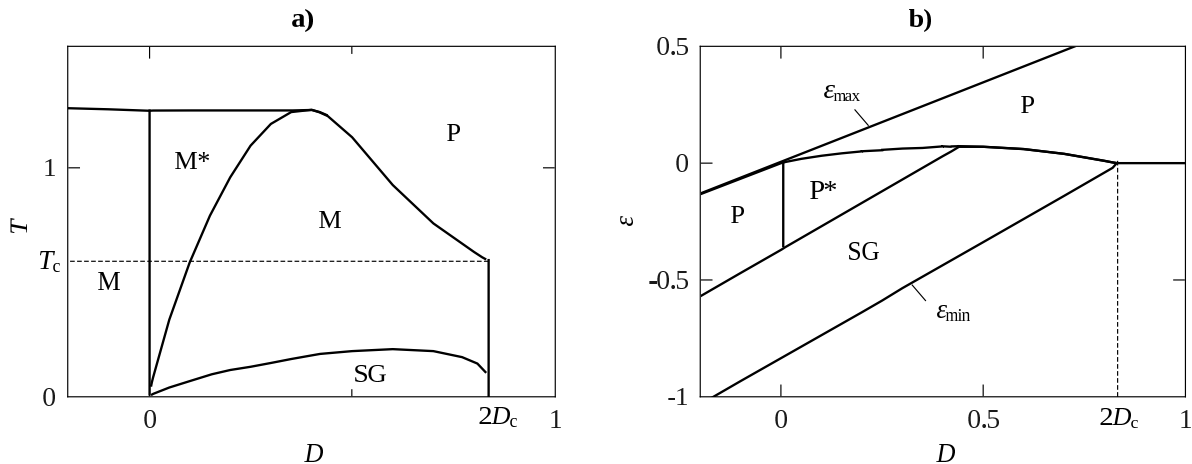} 
\caption{
Microcanonical phase diagrams in the a) $(D,T)$-plane 
and b) $(D,\epsilon)$-plane for $p=37$. 
} 
\label{pgt}
\end{figure}

For $p=3$, the situation is still not very elucidating 
on how the infinite $p$ limit is reached. 
Similarly to the case $p=3$, 
the phase diagram can be obtained numerically for any $p$. 
Contrary to expectations, even $p=37$ shows features 
different from $p\rightarrow\infty$, 
but from figure \ref{pgt} it is clear how that limit is achieved.

As $p$ grows, $D_{\rm MCP}\rightarrow 0$, $D_{\rm s}\rightarrow 0$
and $D_{\rm MC}\to 2 D_{\rm c}$, so that the mixed phase M$^{\rm *}$, 
along with P$^{\rm *}$, disappears completely as $p\rightarrow\infty$. 
Simultaneously, the AT line drifts towards the $\epsilon_{\rm min}$-line, 
so that the 1RSB solution is stable everywhere.
 
For infinite $p$, there is an IP phase, 
characterized by $\chi=1$, which is not present for finite $p$. 
As the phase diagram in the limit, \fref{phase-ct}b), suggests, 
the SG phase off the $\epsilon_{\rm min}$-line, becomes the IP phase.
It is because the SG order parameter satisfies $q<1$ there 
and such solution is inhibited in the limit $p\to\infty$.
We also infer that 
the M phase splits into two parts: P+IP for $T\geq T_{\rm c}$ 
and P+SG for $T< T_{\rm c}$. 

\section{Conclusions}
\label{conclusio}

We have investigated an integer-spin SG model, 
where for finite $p>2$ and a certain, $p$-dependent interval 
of the crystal-field strength $D$, the P-SG transition is of first order. 
This results in ensemble inequivalence along that phase transition. 
While the canonical phase boundary is a single line and 
the phases are cleanly separated in the plane with the temperature 
and the crystal-field strength as axes, 
various mixed phases appear there in the microcanonical analysis. 
This has far reaching consequences for the limit $p\to \infty$, 
which corresponds to the integer-spin analogue of REM. 
There, the ensembles give completely different results 
for finite values of $D$, and coincide only in the Ising limit, 
$D\to -\infty$, as expected. 

To the best of our knowledge, our analysis is the first that 
involves the SG transition and ensemble inequivalence, 
enriching that branch of statistical physics by another important facet. 
Furthermore, our result strengthens the point of view that 
the traditional REM shows ensemble inequivalence within 
the replica ansatz, which have to be contrasted to results that 
do not use the this method.

\section*{Acknowledgments}

We are grateful to Hidetoshi Nishimori for discussions and 
careful reading of the manuscript and Tomoyuki Obuchi 
for useful discussions and comments. 

\appendix

\section{Replica analysis}
\label{details}

Here we give some technical details on 
the derivations of the generalized thermodynamic potentials 
\eref{phi} and \eref{sig}.

\subsection{Canonical ensemble}

Following the standard strategy of the replica method \cite{MPV,Nishimori,MM}, 
we can write the average of the $n$th power of 
the partition function $Z$ in the canonical ensemble as
\be
 [Z^n] 
 &=& \Tr\left\{
 \frac{N\beta^2}{4}
 \sum_{a=1}^n\left(\frac{1}{N}
 \sum_{i=1}^N\left(S_i^a\right)^2 \right)^p 
 +\frac{N\beta^2}{4}
 \sum_{a\ne b} 
 \left(\frac{1}{N}\sum_{i=1}^NS_i^aS_i^b \right)^p
 \right.\no\\
 & & \left.
 -\beta D\sum_{a=1}^n\sum_{i=1}^N\left(S_i^a\right)^2
 \right\}.
\ee 
This expression implies that 
we can introduce two kinds of order parameters 
$\chi_{a}=(1/N)\sum_{i=1}^N(S_i^a)^2$
and $q_{ab}=(1/N)\sum_{i=1}^NS_i^aS_i^b$, with $a\ne b$.
In the present case with integer spins,
$(S_i^a)^2$ is not unity, as in the Ising-spin case and
$1-\chi_a$ measures the fraction of the spins in the zero state.
In the Ising limit $D\to-\infty$, all spins take values $S_i^a=\pm 1$, 
and therefore $\chi_a=1$.
These order parameters are incorporated into the expression of 
the partition function by delta functions.
This generalized Hubbard-Stratonovitch transformation 
gives us the expression
\be
 [Z^n] 
 &=& \int d\chi dq
 \prod_{a}\delta\left(\chi_{a}-\frac{1}{N}\sum_{i}(S_i^a)^2\right) 
 \prod_{a\ne b}\delta\left(q_{ab}-\frac{1}{N}\sum_{i}S_i^aS_i^b\right)
 \no\\
 & & \times\exp\left\{
 \frac{N\beta^2}{4}\sum_{a}(\chi_a)^p
 +\frac{N\beta^2}{2}\sum_{a<b}(q_{ab})^p
 -\beta D\sum_{ai}\left(S_i^a\right)^2\right\}
 \no\\
 &=& \int d\chi dqd\hat{\chi}d\hat{q}
 \exp\left\{
 \frac{N\beta^2}{4}\sum_a(\chi_a)^p 
 +\frac{N\beta^2}{2}\sum_{a<b}(q_{ab})^p
 \right.\no\\
 & & \left.
 -N\sum_{a}\hat{\chi}_a\chi_a 
 -N\sum_{a<b}\hat{q}_{ab}q_{ab} 
 -N\beta D\sum_{a}\chi_a
 +N\ln\Tr e^L
 \right\},
 \label{Zn}
\ee
where 
\be
 L = \sum_{a}\hat{\chi}_{a}S_a^2+\sum_{a<b}\hat{q}_{ab}S_aS_b.
 \label{L}
\ee
The conjugate variables $\hat{q}$ and $\hat{\chi}$ 
are introduced to express the delta functions in their integral forms.
They are related to the order parameters as
\be
 & & \hat{\chi}_a = \frac{\beta^2}{4}p \chi_{a}^{p-1}-\beta D, 
 \label{hatchic}
 \\ 
 & & \hat{q}_{ab} = \frac{\beta^2}{2}p q_{ab}^{p-1}.
 \label{hatqc}
\ee

We impose the 1RSB ansatz for the order parameter $q_{ab}$
and the replica symmetric one for $\chi_a$.
The 1RSB form of the SG order-parameter is 
\be
 \label{qrsb}
 q_{ab} = 
 \left\{\begin{array}{cl}
 0 & \mbox{for $a=b$} \\
 q_1 & \mbox{for
 $\lfloor(a-1)/x\rfloor=\lfloor(b-1)/x\rfloor$} \\
 q_0 & \mbox{otherwise}
 \end{array}\right.,
\ee 
where the Parisi breaking parameter $x$ takes values between 1 and $n$,  
and $\lfloor z \rfloor$ is the floor function giving 
the largest integer not exceeding $z$. 
When spin-reflection symmetry is present, 
$q_0$ is zero and we can set $q_1=q$. 
Then, the spin sum in $\Tr e^L$
can be performed after a Gaussian insertion as follows,
\be
 \Tr e^L &=& \prod_{B=1}^{n/x}\int Dz_{B}
 \Tr\exp\left\{
 \left(\hat{\chi}-\frac{1}{2}\hat{q}\right)\sum_{a=1}^n S_a^2
 +\sqrt{\hat{q}}\sum_{B=1}^{n/x}z_B
 \sum_{a=1+(B-1)x}^{Bx} S_{a}
 \right\} \no\\
 &=& \prod_{B=1}^{n/x}\int Dz_{B}
 \left(
 1+2e^{\hat{\chi}-\hat{q}/2}
 \cosh(\sqrt{\hat{q}}z_B)
 \right)^x \no\\
 &=& \left\{\int Dz
 \left(
 1+2e^{\hat{\chi}-\hat{q}/2}
 \cosh(\sqrt{\hat{q}}z)
 \right)^x\right\}^{n/x},
\ee
where $Dz=dz\exp(-z^2/2)/\sqrt{2\pi}$.
The generalized free energy is defined as 
\be
 \beta\phi=-\lim_{n\to 0}\frac{1}{Nn}\ln[Z^n],
\ee
which then gives \eref{phi}.

\subsection{Microcanonical ensemble}

Next, we consider the microcanonical ensemble.
The density of states for a given energy $E$ is
\be
 \Omega = \Tr\delta(E-H) 
 = \int\frac{dt}{2\pi}\, e^{i(E-H)t}.
\ee
Due to a formal similarity of this expression 
to the canonical partition function,
we can apply the replica analysis to 
calculate the bond-average of the entropy $S=[\ln\Omega]$.
The calculation goes along the same lines 
as in the case of the Ising-spin system~\cite{BN}.
We find the expression 
\be
 [\Omega^n] &=& \exp\left\{
 -N\sum_{ab}
 (\epsilon-D\chi_a)(Q^{-1})_{ab}(\epsilon-D\chi_b)
 -N\sum_a\hat{\chi}_a\chi_a 
 \right.\no\\
 & & \left.
 -N\sum_{a<b}\hat{q}_{ab}q_{ab}+N\ln\Tr e^L\right\},
\ee
 where $\epsilon=E/N$ and $Q$ is an $n\times n$ symmetric matrix 
 whose components are given by
\be
 Q_{ab}=\left\{\begin{array}{cc}
 (\chi_{a})^p & a=b \\
 (q_{ab})^p & a\ne b 
 \end{array}\right..
\ee
 The conjugate variables are obtained from the equations 
\be
 &\hat{\chi}_a 
 = p\sum_{bc}
 \epsilon_b
 (Q^{-1})_{ba}(\chi_{a})^{p-1}
 (Q^{-1})_{ac}\epsilon_c
 +2D\sum_{b}(Q^{-1})_{ab}\epsilon_b,
 \\
 &\hat{q}_{ab} = 
 2p\sum_{cd}
 \epsilon_c
 (Q^{-1})_{ca}
 (q_{ab})^{p-1}
 (Q^{-1})_{bd}
 \epsilon_d, 
\ee
where $\epsilon_a=\epsilon-D\chi_a$

We impose the same 1RSB ansatz as the canonical case 
to the order parameters.
The conjugate variables are given by 
\be
 & & \hat{\chi} = p\chi^{p-1}
 \frac{(\epsilon-D\chi)^2}{\left\{\chi^p-(1-x)q^p\right\}^2}
 +2D\frac{\epsilon-D\chi}{\chi^p-(1-x)q^p}, 
 \label{hatchimc}
 \\
 & & \hat{q} = 2pq^{p-1}
 \frac{(\epsilon-D\chi)^2}{\left\{\chi^p-(1-x)q^p\right\}^2}.
 \label{hatqmc}
\ee
Then, by taking the limit
\be
 \sigma = \lim_{n\to 0}\frac{1}{Nn}\ln[\Omega^n],
\ee
 we obtain \eref{sig}.

\section{Derivation of the AT condition}\label{AT}

\subsection{Expansion around the saddle-point solution}

In this part, 
the AT condition, the stability condition of a certain replica solution, 
is derived.
We use the canonical ensemble to derive the condition, 
which can be easily translated into the microcanonical case.

The starting point is equation \eref{Zn}, or rather its logarithm as
\be
 n\beta\phi &=& -\frac{1}{N}\ln[Z^n] \\
 &=& -\frac{\beta^2}{4}\sum_a(\chi_a)^p 
 -\frac{\beta^2}{2}\sum_{a<b}(q_{ab})^p
 +\sum_{a}\hat{\chi}_a\chi_a 
 +\sum_{a<b}\hat{q}_{ab}q_{ab} 
 +\beta D\sum_{a}\chi_a
 \no\\
 & & 
 -\ln\Tr e^L,
\ee
where $L$ is given in \eref{L}.
Then, we consider variations of the form
\be
 & & \chi_{a}\to \chi_{a}+\delta \chi_{a}, \\
 & & q_{ab}\to q_{ab}+\delta q_{ab},
\ee
and accordingly, $\phi$ is expanded  
as $\phi=f+\delta\phi+\delta_2\phi+\cdots$. 
We shall impose some solution of the free energy 
at the zeroth order of this expansion.
Since the saddle-point method is used 
in the derivation of the free energy, 
the first order of the variation is equal to zero, $\delta\phi=0$. 
The second-order term reads
\be
 \beta n\delta_2\phi &=& 
 \frac{1}{2}\sum_a \delta\chi_a \delta_1\hat\chi_a
 +\frac{1}{2}\sum_{a<b}\delta q_{ab}\delta_1\hat{q}_{ab} \no\\
 & & -\frac{1}{2}\sum_a\sum_b\delta_1\hat{\chi}_a\delta_1\hat{\chi}_b
 \left(\langle S_a^2S_b^2\rangle-\chi_a\chi_b \right) \no\\
 & & -\sum_a\sum_{b<c}\delta_1\hat{\chi}_{a}\delta_1\hat{q}_{bc}
 \left(\langle S_a^2S_bS_c\rangle-\chi_{a}q_{bc} \right) \no\\
 & & -\frac{1}{2}\sum_{a<b}\sum_{c<d}
 \delta_1\hat{q}_{ab}\delta_1\hat{q}_{cd}
 \left(\langle S_aS_bS_cS_d\rangle-q_{ab}q_{cd} \right),
 \label{phi_2}
\ee
where 
\be
 \langle\cdots\rangle = \frac{\Tr \left(\cdots\right)e^L}{\Tr e^L},
\ee
and
\be
 & & \delta_1\hat{\chi}_a 
 = \frac{\beta^2}{4}p(p-1)\chi_a^{p-2}\delta \chi_a, \\
 & & \delta_1\hat{q}_{ab} 
 = \frac{\beta^2}{2}p(p-1)q_{ab}^{p-2}\delta q_{ab}.
 \label{deltahatq}
\ee

Arranging the $n$ elements of $\delta\chi_a$ and the $n(n-1)/2$ elements 
of $\delta q_{ab}$ into a single vector
\be
 \delta\mu^T = (\delta\chi_1,\cdots,\delta\chi_n,
 \delta q_{12},\cdots,dq_{n-1,n}),
\ee
we can define a diagonal matrix $T$ which transforms $\delta\mu$ 
into $\delta\hat{\mu}$, 
which has $\delta\hat{\chi}_a$ and $\delta\hat{q}_{ab}$ as elements:
\be
 \delta\hat{\mu} = T \delta\mu.
\ee
Then, we can write
\be
 \beta n\delta_2\phi 
 = \frac{1}{2}\delta\mu^T(T-TGT)\delta \mu, 
\ee
where the elements of the matrix $G$ can be inferred from \eref{phi_2}.
The stability condition is expressed by equations 
that the eigenvalues of the Hessian matrix $T-TGT$ are nonnegative.

\subsection{Stability of the 1RSB solution}

We impose the replica symmetric solution 
for $\chi_a$ as $\chi_a=\chi$ and 
the 1RSB solution for $q_{ab}$ as \eref{qrsb}
with $q_0=0$, $q_1=q$.

The eigenvalues are calculated 
as the standard case \cite{Nishimori}.
The relevant part of the matrix $G$ is 
components $\delta q_{ab}$ 
with $\lfloor(a-1)/x\rfloor=\lfloor(b-1)/x\rfloor$.
In this block, we have three types of the matrix elements: 
\be
 & & P = G_{(ab)(ab)}=v-q^2, \\
 & & Q = G_{(ab)(ac)}=w-q^2, \\
 & & R = G_{(ab)(cd)}=r-q^2,
\ee
 where the index denoted as $(ab)$ 
 represents the component $\delta q_{ab}$ 
 and different symbols are assumed to be unequal.
 In these expressions,  the spin sums denoted as $v$, $w$ and $r$
 are calculated as follows.
\be
 & & v = \langle (S_{a})^2(S_{b})^2\rangle
 = \int D'z \frac{\left\{2e^{\hat\chi-\hat q/2}
 \cosh(\sqrt{\hat q}z)\right\}^2}
 {\left\{1+2e^{\hat\chi-\hat q/2}\cosh(\sqrt{\hat q}z)\right\}^{2}}, 
 \\
 & & w = \langle S_{a}^2S_{b}S_{c}\rangle
 =  \int D'z \frac{2e^{\hat\chi-\hat q/2}
 \cosh(\sqrt{\hat q}z)
 \left\{2e^{\hat\chi-\hat q/2}\sinh(\sqrt{\hat q}z)\right\}^2}
 {\left\{1+2e^{\hat\chi-\hat q/2} \cosh(\sqrt{\hat q}z)\right\}^{3}}, 
 \\
 & & r = \langle S_{a}S_{b}S_{c}S_{d}\rangle
 =  \int D'z \frac{\left\{2e^{\hat\chi-\hat q/2}
 \sinh(\sqrt{\hat q}z)\right\}^4}
 {\left\{1+2e^{\hat\chi-\hat q/2} \cosh(\sqrt{\hat q}z)\right\}^{4}}.
\ee
 Here, the integral measure is defined as in equation \eref{Dz}.

The stability condition is analogous to the half-integer case, 
where only one eigenvalue of the Hessian can be negative. 
The other eigenvalues correspond to extremization conditions of 
the free energy and are, therefore, positive by construction. 
We have the stability condition
\be
 \frac{2}{\beta^2 p(p-1)q^{p-2}}-(P-2Q+R)\ge 0,
\ee
 where the first term on the left hand side 
 comes from the inverse of the corresponding component of $T$ 
 and can be extracted from \eref{deltahatq}.
 We finally obtain the AT condition in the canonical ensemble
\be
 \frac{2}{p(p-1)\beta^2q^{p-2}} \ge
 \int D'z
 \frac{\left\{
 (2e^{\tilde{\chi}-\tilde{q}/2})^2
 +2e^{\tilde{\chi}-\tilde{q}/2}\cosh(\sqrt{\tilde{q}}z)
 \right\}^2}
 {\left\{1+2e^{\tilde{\chi}-\tilde{q}/2}
 \cosh(\sqrt{\tilde{q}}z)\right\}^4}, 
\ee
where the conjugate variables are given by \eref{hatchic} and \eref{hatqc}.

It is straightforward to convert the canonical result 
into the microcanonical one using the relation \eref{caloric}, 
\be
 \frac{\left\{\chi^p-(1-x)q^p\right\}^2}
 {2p(p-1)q^{p-2}(\epsilon-D\chi)^2} \ge
 \int D'z
 \frac{\left\{
 (2e^{\tilde{\chi}-\tilde{q}/2})^2
 +2e^{\tilde{\chi}-\tilde{q}/2}\cosh(\sqrt{\tilde{q}}z)
 \right\}^2}
 {\left\{1+2e^{\tilde{\chi}-\tilde{q}/2}
 \cosh(\sqrt{\tilde{q}}z)\right\}^4}, \no\\
\ee
where the conjugate variables are given by \eref{hatchimc} and \eref{hatqmc}.


\end{document}